\documentclass[aps,apl,twocolumn,groupedaddress,amsmath]{revtex4}
\usepackage[latin1]{inputenc}    % Accept european-encoded (latin1) characters.
\usepackage{graphicx}   % For eps figures
\usepackage{xspace}
\usepackage{color}

\begin{document}

% some macros
%-----------------------------------------------
%\def\Tone{$\mathrm{T_1}$}
%\def\Ttwo{$\mathrm{T_2}$}
%\def\TtwoE{$\mathrm{T^E_2}$}
%\def\Tph{$\mathrm{T^E_{\phi}}$}
%\newcommand{\mpar}[1]{\marginpar{\small \textcolor{magenta}{#1}}}
\newcommand{\mpar}[1]{\marginpar{\small \it \textcolor{magenta}{#1}}}

\newcommand*{\Hint}{\ensuremath{\hat{H}_\mathrm{int}}\xspace}
\newcommand*{\Hsub}{\ensuremath{\hat{H}_\mathrm{sub}}\xspace}
\newcommand*{\Hqb}{\ensuremath{\hat{H}_\mathrm{qb}}\xspace}
\newcommand*{\Htls}{\ensuremath{\hat{H}_\mathrm{TLS}}\xspace}
\newcommand*{\siX}{\ensuremath{\hat{\sigma}_\mathrm{x}}\xspace}
\newcommand*{\siZ}{\ensuremath{\hat{\sigma}_\mathrm{z}}\xspace}
\newcommand*{\siXq}{\ensuremath{\hat{\sigma}_\mathrm{x}^\mathrm{qb}}\xspace}
\newcommand*{\siYq}{\ensuremath{\hat{\sigma}_\mathrm{y}^\mathrm{qb}}\xspace}
\newcommand*{\siZq}{\ensuremath{\hat{\sigma}_\mathrm{z}^\mathrm{qb}}\xspace}
\newcommand*{\siXt}{\ensuremath{\hat{\sigma}_\mathrm{x}^\mathrm{TLS}}\xspace}
\newcommand*{\siYt}{\ensuremath{\hat{\sigma}_\mathrm{y}^\mathrm{TLS}}\xspace}
\newcommand*{\siZt}{\ensuremath{\hat{\sigma}_\mathrm{z}^\mathrm{TLS}}\xspace}
\newcommand*{\siXs}{\ensuremath{\hat{\sigma}_\mathrm{x}^\mathrm{sub}}\xspace}
\newcommand*{\siYs}{\ensuremath{\hat{\sigma}_\mathrm{y}^\mathrm{sub}}\xspace}
\newcommand*{\siZs}{\ensuremath{\hat{\sigma}_\mathrm{z}^\mathrm{sub}}\xspace}
\newcommand*{\Isq}{\ensuremath{I_\mathrm{b}}\xspace}
\newcommand*{\PhiQ}{\ensuremath{\Phi_\mathrm{qb}}\xspace}
\newcommand*{\fqb}{\ensuremath{f_\mathrm{qb}}\xspace}
\newcommand*{\ftls}{\ensuremath{f_\mathrm{TLS}}\xspace}
\newcommand*{\fosc}{\ensuremath{f_\mathrm{osc}}\xspace}
\newcommand*{\dph}{\ensuremath{\delta\Phi}\xspace}
\newcommand*{\df}{\ensuremath{\delta\!f}\xspace}
\newcommand*{\dpint}{\ensuremath{\delta\Phi_\mathrm{int}}\xspace}
\newcommand*{\TphN}{\ensuremath{T_{\varphi,\mathrm{N}}}\xspace}
\newcommand*{\Tph}[1]{\ensuremath{T_{\varphi,\mathrm{#1}}}\xspace}
\newcommand*{\tp}{\ensuremath{t_\mathrm{p}}\xspace}
\newcommand*{\fRabi}{\ensuremath{f_\mathrm{Rabi}}\xspace}
\newcommand*{\Ian}{\ensuremath{I_\mathrm{antenna}^\mathrm{mw}}\xspace}
\newcommand*{\Imr}{\ensuremath{I_\mathrm{r}^\mathrm{mw}}\xspace}
\newcommand*{\emd}{\ensuremath{\varepsilon^\mathrm{mw}_\mathrm{direct}}\xspace}
\newcommand*{\emt}{\ensuremath{\varepsilon^\mathrm{mw}}\xspace}

\newcommand{\ket}[1]{\vert  #1 \rangle} 
\newcommand{\bra}[1]{\langle  #1 \vert}

\newcommand*{\PhiX}{\ensuremath{\Phi_\mathrm{X}}\xspace}
\newcommand*{\PhiZ}{\ensuremath{\Phi_\mathrm{Z}}\xspace}
\newcommand*{\fX}{\ensuremath{f_\mathrm{x}}\xspace}
\newcommand*{\fZ}{\ensuremath{f_\mathrm{z}}\xspace}
\newcommand*{\Ax}{\ensuremath{A_\mathrm{x}}\xspace}
\newcommand*{\Az}{\ensuremath{A_\mathrm{z}}\xspace}

\newcommand*{\TF}{\ensuremath{T_{\varphi F}}\xspace}
\newcommand*{\TE}{\ensuremath{T_{\varphi E}}\xspace}
\newcommand*{\GF}{\ensuremath{\Gamma_{\varphi F}}\xspace}
\newcommand*{\GE}{\ensuremath{\Gamma_{\varphi E}}\xspace}

\newcommand*{\mPh}{\ensuremath{\,\mathrm{m}\Phi_0}\xspace}
\newcommand*{\uPh}{\ensuremath{\,\mu\Phi_0}\xspace}

\newcommand*{\um}{\ensuremath{\,\mu\mathrm{m}}\xspace}
\newcommand*{\nm}{\ensuremath{\,\mathrm{nm}}\xspace}
\newcommand*{\mm}{\ensuremath{\,\mathrm{mm}}\xspace}
\newcommand*{\m}{\ensuremath{\,\mathrm{m}}\xspace}
\newcommand*{\sqm}{\ensuremath{\,\mathrm{m}^2}\xspace}
\newcommand*{\sqmm}{\ensuremath{\,\mathrm{mm}^2}\xspace}
\newcommand*{\squm}{\ensuremath{\,\mu\mathrm{m}^2}\xspace}
\newcommand*{\psqm}{\ensuremath{\,\mathrm{m}^{-2}}\xspace}
\newcommand*{\psqmV}{\ensuremath{\,\mathrm{m}^{-2}\mathrm{V}^{-1}}\xspace}
\newcommand*{\cm}{\ensuremath{\,\mathrm{cm}}\xspace}

\newcommand*{\nF}{\ensuremath{\,\mathrm{nF}}\xspace}
\newcommand*{\pF}{\ensuremath{\,\mathrm{pF}}\xspace}
\newcommand*{\pH}{\ensuremath{\,\mathrm{pH}}\xspace}

\newcommand*{\emob}{\ensuremath{\,\mathrm{m}^2/\mathrm{V}\mathrm{s}}\xspace}
\newcommand*{\edos}{\ensuremath{\,\mu\mathrm{C}/\mathrm{cm}^2}\xspace}
\newcommand*{\mbar}{\ensuremath{\,\mathrm{mbar}}\xspace}

\newcommand*{\A}{\ensuremath{\,\mathrm{A}}\xspace}
\newcommand*{\mA}{\ensuremath{\,\mathrm{mA}}\xspace}
\newcommand*{\nA}{\ensuremath{\,\mathrm{nA}}\xspace}
\newcommand*{\pA}{\ensuremath{\,\mathrm{pA}}\xspace}
\newcommand*{\fA}{\ensuremath{\,\mathrm{fA}}\xspace}
\newcommand*{\uA}{\ensuremath{\,\mu\mathrm{A}}\xspace}

\newcommand*{\Ohm}{\ensuremath{\,\Omega}\xspace}
\newcommand*{\kOhm}{\ensuremath{\,\mathrm{k}\Omega}\xspace}
\newcommand*{\MOhm}{\ensuremath{\,\mathrm{M}\Omega}\xspace}
\newcommand*{\GOhm}{\ensuremath{\,\mathrm{G}\Omega}\xspace}

\newcommand*{\Hz}{\ensuremath{\,\mathrm{Hz}}\xspace}
\newcommand*{\kHz}{\ensuremath{\,\mathrm{kHz}}\xspace}
\newcommand*{\MHz}{\ensuremath{\,\mathrm{MHz}}\xspace}
\newcommand*{\GHz}{\ensuremath{\,\mathrm{GHz}}\xspace}
\newcommand*{\THz}{\ensuremath{\,\mathrm{THz}}\xspace}

\newcommand*{\K}{\ensuremath{\,\mathrm{K}}\xspace}
\newcommand*{\mK}{\ensuremath{\,\mathrm{mK}}\xspace}

\newcommand*{\kV}{\ensuremath{\,\mathrm{kV}}\xspace}
\newcommand*{\V}{\ensuremath{\,\mathrm{V}}\xspace}
\newcommand*{\mV}{\ensuremath{\,\mathrm{mV}}\xspace}
\newcommand*{\uV}{\ensuremath{\,\mu\mathrm{V}}\xspace}
\newcommand*{\nV}{\ensuremath{\,\mathrm{nV}}\xspace}

\newcommand*{\eV}{\ensuremath{\,\mathrm{eV}}\xspace}
\newcommand*{\meV}{\ensuremath{\,\mathrm{meV}}\xspace}
\newcommand*{\ueV}{\ensuremath{\,\mu\mathrm{eV}}\xspace}

\newcommand*{\T}{\ensuremath{\,\mathrm{T}}\xspace}
\newcommand*{\mT}{\ensuremath{\,\mathrm{mT}}\xspace}
\newcommand*{\uT}{\ensuremath{\,\mu\mathrm{T}}\xspace}

\newcommand*{\ms}{\ensuremath{\,\mathrm{ms}}\xspace}
\newcommand*{\s}{\ensuremath{\,\mathrm{s}}\xspace}
\newcommand*{\us}{\ensuremath{\,\mathrm{\mu s}}\xspace}
\newcommand*{\ns}{\ensuremath{\,\mathrm{ns}}\xspace}
\newcommand*{\rpm}{\ensuremath{\,\mathrm{rpm}}\xspace}
\newcommand*{\minute}{\ensuremath{\,\mathrm{min}}\xspace}
\newcommand*{\degree}{\ensuremath{\,^\circ\mathrm{C}}\xspace}

\newcommand*{\EqRef}[1]{Eq.\,(\ref{#1})}
\newcommand*{\FigRef}[1]{Fig.\,\ref{#1}}
\newcommand*{\dd}[2]{\mathrm{\partial}#1/\mathrm{\partial}#2}
\newcommand*{\ddf}[2]{\frac{\mathrm{\partial}#1}{\mathrm{\partial}#2}}

\title{Dynamical decoupling and dephasing in interacting two-level systems}
 \author{Simon Gustavsson$^1$}
% \email{simongus@mit.edu}
 \author{Fei Yan$^2$}
 \author{Jonas Bylander$^1$}
 \author{Fumiki Yoshihara$^3$}
% \author{Khalil Harrabi$^{3,\dag}$}
 \author{Yasunobu Nakamura$^{3,4,\ddag}$}
 \author{Terry P. Orlando$^{1}$}
 \author{William D. Oliver$^{1,5}$}
 \affiliation{$^1$Research Laboratory of Electronics, Massachusetts Institute of Technology, Cambridge, MA 02139, USA \\
  $^2$Department of Nuclear Science and Engineering, MIT, Cambridge, MA 02139, USA \\
  $^3$The Institute of Physical and Chemical Research (RIKEN), Wako, Saitama 351-0198, Japan \\
  $^4$Green Innovation Research Laboratories, NEC Corporation, Tsukuba, Ibaraki 305-8501, Japan\\
  $^5$MIT Lincoln Laboratory, 244 Wood Street, Lexington, MA 02420, USA \\
%  $^\dag$Present address:  Physics Department, King Fahd University of Petroleum \& Minerals, Dhahran 31261, Saudi Arabia\\
  $^\ddag$Present address:  Research Center for Advanced Science and Technology (RCAST), University of Tokyo, Komaba, Meguro-ku, Tokyo 153-8904, Japan}

%\date{\today}
\begin{abstract}
% what we did (DD, mitigate noise, coupling
We implement dynamical decoupling techniques to mitigate noise and enhance the lifetime of an entangled state that is formed in a superconducting flux qubit coupled to a microscopic two-level system.
% performance
By rapidly changing the qubit's transition frequency relative to the two-level system, we realize a refocusing pulse that reduces dephasing due to fluctuations in the transition frequencies, thereby improving the coherence time of the entangled state.
% multi-pulse
The coupling coherence is further enhanced when applying multiple refocusing pulses, in agreement with our $1/f$ noise model. 
The results are applicable to any two-qubit system with transverse coupling, and they highlight the potential of decoupling techniques for improving two-qubit gate fidelities, an essential prerequisite for implementing fault-tolerant quantum computing.
%% outlook
%The results highlight the potential of decoupling techniques for improving two-qubit gate fidelities, an essential prerequisite for implementing fault-tolerant quantum computing.
\end{abstract}

\maketitle
% introduction
% - two qubit gates for quantum computing
%Any two-qubit entangling gate is, together with single-qubit rotations, a sufficient requirement for implementing universal quantum algorithms \cite{Bremner:2002}.
%
A universal set of quantum gates, sufficient for implementing any quantum algorithm, consists of a two-qubit entangling gate together with single-qubit rotations \cite{Bremner:2002}.
% - requirements for gate fidelity
However, fault-tolerant quantum computing with error-correcting protocols sets strict limits on the allowable error rate of each gate. %, signifying the importance of improving gate fidelities.
% - improvement on gates
Initial work focused on perfecting single-qubit gates \cite{Knill:2008, Lucero:2008, Chow:2010}, but in recent years there has been progress on characterizing two-qubit gate operations \cite{Bialczak:2010,Chow:2011,Dewes:2012}.
% - previous work on coupled qubits, TLS
In superconducting systems, two-qubit gates have been implemented in a variety of ways, for example through geometric couplings \cite{Yamamoto:2003, Steffen:2006}, tunable coupling elements \cite{Hime:2006, Bialczak:2011}, microwave resonators \cite{Majer:2007,Sillanpaa:2007}, or with microwave-induced interactions \cite{Plantenberg:2007, Niskanen:2007, Rigetti:2010, Groot:2010, Chow:2011}.
% - noisy coupling parameter, regardless of coupling method
Regardless of the nature of the coupling, any variations of the qubit frequencies or in the coupling parameter during the two-qubit interaction leads to dephasing of the entangled state, and puts an upper limit on the obtainable gate fidelity \cite{Bialczak:2010}.

% our work
% - spin echo, single qubits
For single qubits, dephasing due to low-frequency fluctuations in the precession frequency is routinely reduced with refocusing techniques \cite{Chiorescu:2003,Petta:2005,Ithier:2005}, originally developed in nuclear magnetic resonance \cite{Hahn:1950}. 
% - refocusing of noise in the coupling
In this work, we apply similar techniques to improve the coherence of an entangled state formed between a flux qubit and a microscopic two-level system (TLS).
% - implementation
% -- something about decaying swaps
The refocusing pulse is implemented by rapidly changing the qubit frequency relative to the TLS, thereby acquiring a phase shift \cite{Hofheinz:2009}.
When the phase shift equals $\pi$, the pulse refocuses the incoherent evolution of the coupled qubit-TLS system, giving a fourfold improvement of its coherence time.
%For properly calibrated pulses, this will refocus dephasing due to incoherent broadening of the coupling parameter, giving a fourfold improvement of the coupling coherence time.
% - extending dynamical decoupling
We further prolong the decay times by applying multiple refocusing pulses \cite{Carr:1954, Bylander:2011}, thus extending dynamical-decoupling techniques \cite{Viola:1999} to correct for dephasing of entangled states.
% - first step towards optimal control
The results are first steps towards implementing error-correcting composite gate pulses \cite{Cummins:2003, Kerman:2008} and optimal control methods \cite{Khaneja:2005}, schemes with strong potential for improving two-qubit gate operations. %, regardless of the physical coupling method.

\begin{figure}[t!]
\centering
\includegraphics[width=\linewidth]{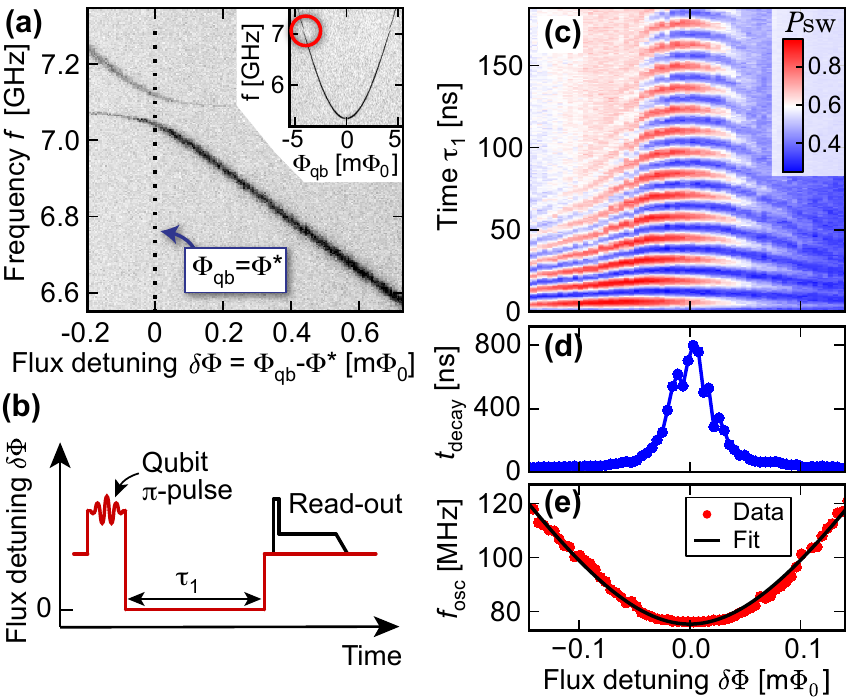}
\caption{(a) Spectroscopy of the qubit-TLS system.  The qubit and TLS are resonant at $f=7.08\GHz$, where the spectrum has an anticrossing with splitting $S=76\MHz$. The inset shows the qubit spectrum over a larger range, with the red circle indicating the region of interest.
(b) Pulse sequence for probing the qubit-TLS interactions. The $\pi$-pulse generates a qubit excitation, which is coherently exchanged back and forth between qubit and TLS during the interaction time $\tau_1$. 
(c) Coherent oscillations between qubit and TLS, measured using the pulse sequence shown in (b). High switching probability $P_{SW}$ corresponds to the qubit's ground state $\ket{0}$, low $P_{SW}$ to the qubit's excited state $\ket{1}$.
%For each value of $\dph$, we have subtracted the average value of $P_\mathrm{SW}$ over $\tau_1$ to improve the visibility of the oscillations.
(d-e) Characteristic decay time $t_\mathrm{decay}$ and oscillation frequency $\fosc$, extracted from the data in (c).  The oscillations decay faster for $\dph \neq 0$, a consequence of the increased sensitivity $\dd{\fosc}{\Phi}$ to flux noise. 
}
\label{fig:Setup}
\end{figure}

% sample
We use a flux qubit \cite{Mooij:1999}, consisting of a superconducting loop interrupted by four Josephson junctions (see Ref.\,\cite{Bylander:2011} for a detailed description of the device).  The qubit's diabatic states correspond to clockwise and counterclockwise persistent currents $\pm I_\mathrm{P}$, with $I_\mathrm{P}=180\nA$.  The inset of \FigRef{fig:Setup}(a) shows a spectrum of the device versus external flux, with $\PhiQ$ defined as $\PhiQ = \Phi + \Phi_0/2$ and $\Phi_0= h/2e$.  The qubit frequency follows $\fqb = \sqrt{\Delta^2 + \varepsilon^2}$, where the tunnel coupling $\Delta = 5.4\GHz$ is set by the design parameters and the energy detuning $\varepsilon = 2 I_\mathrm{P} \PhiQ/h$ is controlled by the applied flux.  The device is embedded in a SQUID, which is used as a sensitive magnetometer for qubit read-out \cite{Chiorescu:2003}.

% TLS
At $\PhiQ\,$=$\,\Phi^*\,$=$\,\pm4.15\mPh$, the qubit becomes resonant with a TLS \cite{Simmonds:2004}.  The microscopic nature of the TLS is unknown, but studies of two-level systems in similar qubit designs show that the most likely origin is an electric dipole in one of the tunnel junctions \cite{Lupascu:2009}.
Figure \ref{fig:Setup}(a) shows a magnification of the region around $-4.15\mPh$, revealing a clear anticrossing with splitting $S = 76 \MHz$.  We describe the  system using the four states $\{\ket{0g},\,\ket{1g},\,\ket{0e},\,\ket{1e}\}$, where $(0,\,1)$ are the qubit energy eigenstates and $(g,\,e)$ refer to the ground and excited state of the TLS.  On resonance, $\ket{1g}$ and $\ket{0e}$ are degenerate and coupled by the coupling energy $hS$.
% - method for investigating
To characterize the coupling, we use the pulse scheme depicted in \FigRef{fig:Setup}(b) \cite{Neeley:2008}.  Starting with both qubit and TLS in their ground states $(\ket{0g})$, we rapidly shift the flux to a position $\dph = \PhiQ  - \Phi^* = 1.2\mPh$ where the qubit frequency $\fqb = 6.3\GHz$ is far detuned from the TLS.  By applying a microwave pulse, resonant with $\fqb$, we perform a $\pi$-rotation on the qubit and put the system in $\ket{1g}$.   We then rapidly shift $\dph$ to a value close to zero, effectively turning on the interaction $S$, whereupon the system will oscillate between $\ket{1g}$ and $\ket{0e}$.  After a time $\tau_1$, the interaction is turned off by shifting $\dph$ away from zero, and we measure the final qubit state by applying a read-out pulse to the SQUID.  Since the measurement outcome is stochastic, we repeat the sequence a few thousand times to acquire sufficient statistics to estimate the SQUID switching probability $P_\mathrm{SW}$ and thereby the qubit state.
%We repeat the sequence a few thousand times to determine the SQUID switching probability $P_\mathrm{SW}$.

% - oscillations
Figure \ref{fig:Setup}(c) shows the qubit state after the pulse sequence, measured versus interaction time $\tau_1$ and flux detuning $\dph$.  At $\dph=0$ and for $\tau_1 = 1/(2S) = 6\ns$, the pulse sequence implements an iSWAP gate between qubit and TLS, taking $\ket{0e} \rightarrow i\ket{1g}$ and $\ket{1g} \rightarrow i\ket{0e}$ \cite{Schuch:2003, Zagoskin:2006, Neeley:2008}. 
%Figures \ref{fig:Setup}(d-e) show the characteristic decay time and the frequency $\fosc$ of the oscillations. 
The characteristic decay time of the oscillations is shown in \FigRef{fig:Setup}(d). 
%The oscillations frequency $\fosc$ increases for $\dph \neq 0$, due to frequency detuning between the qubit and the TLS.
The oscillations persist the longest at $\dph=0$; at this point, the decay time is $\sim\!\!800\ns$.
% and mainly limited by energy relaxation to $\ket{0g}$. 
% - frequency and decay
However, as $\dph$ is moved away from the optimal point, the decay time quickly decreases to zero.
% describe reason for dephasing
We attribute the reduction in coupling coherence to low-frequency flux noise, present in all superconducting devices \cite{Wellstood:1987}. 
% - flux noise couples to oscillation frequency
When $\dph \neq 0$, fluctuations in $\dph$ induce variations in the effective coupling frequency $\fosc$ [\FigRef{fig:Setup}(e)], leading to dephasing of the entangled state. 
%At $\dph=0$, $\fosc$ is first-order insensitive to changes in $\dph$, and the system is effectively decoupled from flux noise.
% dynamical decupling
%A different approach to reducing the influence of noise is to perform dynamical decoupling, where a sequence of rotations

% - normally refocused with spin-echo
For single qubits, dephasing due to low-frequency fluctuations of the qubit frequency can be reduced in a Hahn-echo experiment \cite{Hahn:1950}.  By applying a $\pi$-pulse after a time $t$ of dephasing, the qubit's noise-induced evolution will reverse directions and refocus at time $2t$, provided that the fluctuations are slow on the time scale $2t$ \cite{Biercuk:2009, Bylander:2011}.  
% way to apply refocusing pulses
Here, our goal is to extend such single-qubit refocusing techniques to the mitigation of noise in coupled systems with multiple qubits, which requires implementing refocusing pulses for entangled states.  Note that the purpose here is to increase coherence times, as opposed to turning off unwanted couplings \cite{Leung:2000, Vandersypen:2001}.

\begin{figure}[tb]
\centering
\includegraphics[width=\linewidth]{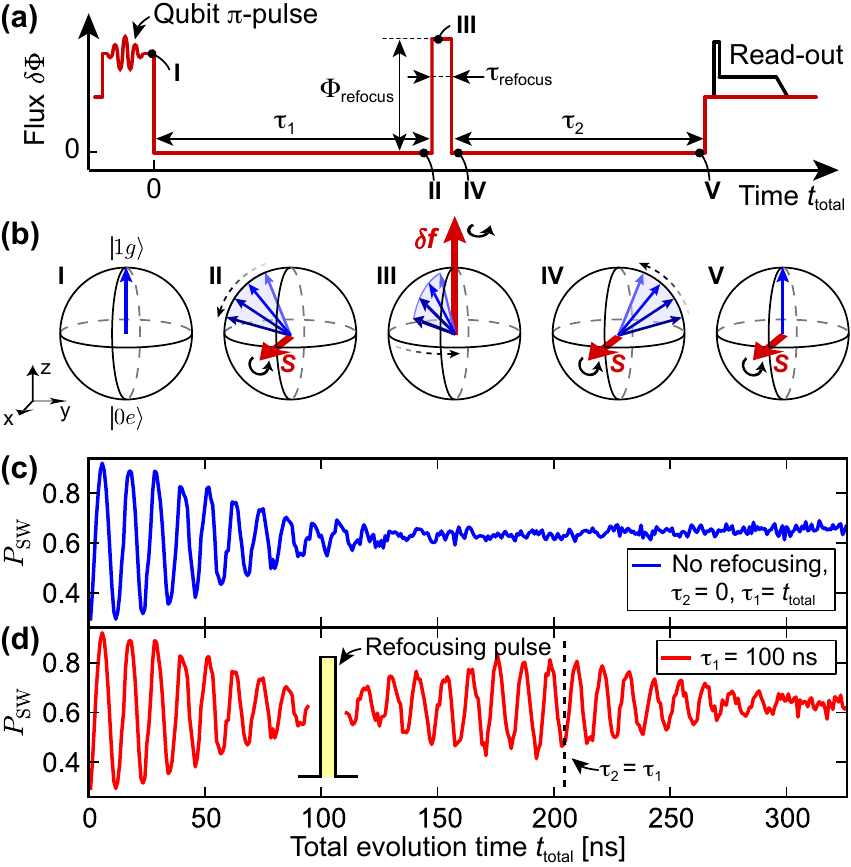}
\caption{(a) Pulse sequence and (b) Bloch sphere representation of the refocusing protocol. The blue arrows are state vectors, while the red arrows represent the Hamiltonian in \EqRef{eq:Hsub}.
 After the qubit $\pi$ pulse, the system enters the $\{\ket{1g},\ket{0e}\}$ subspace (step I).  The coupling $S$ rotates the state  vector around the $x$-axis, but due to noise in the effective coupling the state vector fans out (II).  The rapid flux pulse $\Phi_\mathrm{refocus}$ generates a large frequency detuning $\df$, the system will start rotating around the $z$-axis (III) and eventually complete a $\pi$-rotation (IV). The inhomogeneous broadening now refocuses the state vector, giving an echo at V.
 (c-d) Evolution of the qubit-TLS system, measured with and without a refocusing pulse.  A clear echo appears after the refocusing pulse, with a maximum close to $\tau_2 = \tau_1$. The traces were taken at $\dph = -72\uPh$.}
\label{fig:Bloch}
\end{figure}

% which requires a method to implement refocusing pulses in the coupled system.
%To show how such refocusing can be realized, 
We start by describing the system's dynamics. 
% hamiltonian
Following Refs.\,\cite{Simmonds:2004, Zagoskin:2006, Lupascu:2009}, we write the total Hamiltonian as 
$\hat{H} = \Hqb + \Htls + \Hint$, with $\Hqb = - (h/2) \fqb \, \siZq$, $\Htls = - (h/2) \ftls \, \siZt$, and with the interactions described by $\Hint = -(h/2) S\, \siXq \siXt$.  Here, 
%\begin{equation}
%  \Hint = -\frac{h}{2} S \, \siXq \siXt. \label{eq:Hint} 
%\end{equation}
$\hat{\sigma}_\mathrm{x,z}^\mathrm{qb}$ are Pauli operators for the qubit, $\hat{\sigma}_\mathrm{x,z}^\mathrm{TLS}$ are TLS operators and $\ftls$ is the TLS frequency.   To focus on the interactions between the qubit and the TLS, we restrict the discussion to the subspace spanned by the states $\{\ket{1g},\,\ket{0e}\}$.  In the rotating frame of the TLS, the subspace Hamiltonian becomes
\begin{equation}
  \Hsub =  - \frac{h}{2} \left( \df \, \siZs + S \, \siXs \right), \label{eq:Hsub} 
\end{equation}
%\begin{equation}
%  \Hsub =  - \frac{h \df}{2} \siZs - \frac{hS}{2} \siXs, \label{eq:Hsub} 
%\end{equation}
where $\df = \ftls-\fqb$ and $\hat{\sigma}_\mathrm{x,z}^\mathrm{sub}$ are subspace Pauli operators.  The dynamics of \EqRef{eq:Hsub} can be visualized on a Bloch sphere, with the north and south poles corresponding to $\ket{1g}$ and $\ket{0e}$, respectively, and with $S$ and $\df$ representing the length of torque vectors along the $x$- and $z$-axes [see \FigRef{fig:Bloch}(b)].  The frequency of the coherent oscillations seen in \FigRef{fig:Setup}(c) is then given by the effective coupling strength %, $\fosc=  \sqrt{\df^2 + S^2}$,
%The Hamiltonian in \EqRef{eq:Hsub} is easily diagonalized, giving an expression for the frequency $\fosc$  of the coherent oscillations seen in \FigRef{fig:Setup}(c):
\begin{equation}
  \fosc=  \sqrt{\df^2 + S^2}, \label{eq:fosc} 
\end{equation}
which is plotted together with the data in \FigRef{fig:Setup}(e).

% refocusing sequence
With the coupling dynamics described by \EqRef{eq:Hsub}, we discuss the details of the refocusing sequence, shown in  \FigRef{fig:Bloch}(a-b).  The system is brought into the $\{\ket{1g},\,\ket{0e}\}$ subspace by applying a $\pi$-pulse to the qubit [step I in Figs.\,\ref{fig:Bloch}(a-b)], followed by a non-adiabatic shift in $\dph$ to bring the qubit and  TLS close to resonance.  $\ket{1g}$ is not an eigenstate of the coupled system, so the interaction $S$ will cause the system to rotate around the $x$-axis, oscillating between $\ket{1g}$ and $\ket{0e}$.  Low-frequency fluctuations in the effective coupling strength will cause the Bloch state vector to fan out (over many realizations of the experiment), and the system loses its phase coherence (step II).  
%With the coupling dynamics described by \EqRef{eq:Hsub}, we proceed with discussing the refocusing sequence, shown in  \FigRef{fig:Bloch}(a-b).  First, a $\pi$ pulse is applied to the qubit, taking the system into the $\{\ket{1g},\,\ket{0e}\}$ subspace [case I in Figs.\,\ref{fig:Bloch}(a-b)].  We have $\df^2 \gg S^2$, the Bloch vector is close to the $\ket{1g}$ eigenstate, and the coupling is effectively turned off.  We then

% refocusing pulse
% $\dph = \Phi_\mathrm{refocus}$
The refocusing pulse is now implemented by applying a flux shift pulse that rapidly detunes the qubit and the TLS to  $\df=550\MHz$.  With $|\df| \gg |S|$, the state vector is effectively rotating around the $z$-axis (step III) \cite{Petta:2005}, and we realize a $\pi$ rotation by setting the pulse duration $\tau_\mathrm{refocus} = 0.5/\df$.  The system is then rapidly brought back into resonance (step IV), and the state vector continues to rotate around the $x$-axis.  The inhomogeneous broadening that caused the state vector to diffuse during the first interval $\tau_1$ will now realign them again. The refocusing is complete after a time $\tau_2=\tau_1$ (step V). %, at which point we read out the qubit state. %Because of the $\pi$-pulse around the $z$-axis, t
% examples
Figures~\ref{fig:Bloch}(c-d) illustrate the result of the refocusing sequence.  Without the refocusing pulse [\FigRef{fig:Bloch}(c)], the coherent oscillations between $\ket{1g}$ and $\ket{0e}$ decay almost completely after $100\ns$.   When inserting a refocusing pulse at $\tau_1 = 100\ns$, the oscillations start to revive, eventually forming an echo at $\tau_2 = \tau_1$. %, and then decay again as $\tau_2$ is increased further. 
\begin{figure}[tb]
\centering
\includegraphics[width=\linewidth]{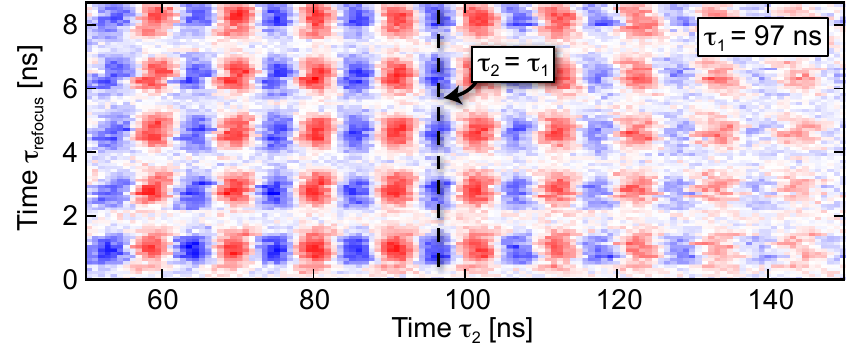}
\caption{ Calibration of the refocusing pulse, measured by fixing $\tau_1 = 97 \ns$ and looking for the echo signal by sweeping $\tau_2$. The echo appears every time the refocusing pulse generates a rotation by an odd integer of $\pi$.  The first five refocusing conditions occur at $\tau_\mathrm{refocus} = (2n+1)\times 0.5/\df  = 0.9,\,2.7,\,4.5,\, 6.4,\,\mathrm{and}\,8.2\ns$, with $\df = 550\MHz$.
}
\label{fig:Calib}
\end{figure}

% calibration
The revival of phase coherence seen in \FigRef{fig:Bloch}(d) requires careful calibration of the refocusing pulse.  Figure~\ref{fig:Calib} shows an example of a calibration experiment, where we fix $\tau_1=97\ns$ and $\df = 550\MHz$ and measure refocused oscillations versus the refocusing time $\tau_\mathrm{refocus}$.  The data shows strong oscillations whenever the  refocusing pulse rotates the state vector by an odd integer of $\pi$, i.e. when $\tau_\mathrm{refocus} = (2n+1)\times 0.5/\df$, in agreement with the schematics discussed in \FigRef{fig:Bloch}(b).  
% = 0.9,\,2.7,\,4.5,\, 6.4,\,\mathrm{and}\,8.2\ns
%In the experiments, we typically choose to operate with $\tau_\mathrm{refocus} $

% fit function
We now turn to investigating the decoherence mechanisms and determining the performance of the refocusing protocol.  For Gaussian-distributed dephasing noise, we expect the amplitude $h(t)$ of the coherent oscillations to decay as \cite{Falci:2005,Ithier:2005}
\begin{equation}
%  p(t) = e^{-\Gamma_1 t} \, e^{-(\Gamma_\varphi^\mathrm{N} t)^2}. \label{eq:Decay} 
  h(t) = \mathrm{exp}[{-t/\tilde{T_1}}] \, \mathrm{exp}[-(t/\TphN)^2]. \label{eq:Decay} 
%  h(t) = \mathrm{exp}[{-\Gamma_1 t}] \, \mathrm{exp}[-(\Gamma_\varphi^\mathrm{N} t)^2]. \label{eq:Decay} 
\end{equation}
The exponential decay constant $\tilde{T_1}$ is due to energy relaxation, while $\TphN$ represents the dephasing with $N$ refocusing pulses.
At $\dph=0$, we measure a pure exponential decay with time constant $\tilde{T_1} = 800\ns$, which is shorter than the relaxation time of both the qubit ($T_1^\mathrm{qb} = 10 \us$) and the TLS ($T_1^\mathrm{TLS} = 1\us$). However, to get an expression for $\tilde{T_1}$, we need to consider all the possible absorption/emission rates in the full four-level system \cite{Paladino:2010}. In the relevant situation $hS \ll k_\mathrm{B} T \ll h\ftls,\, h\fqb$, we have
\begin{equation}
1/\tilde{T_1} = \frac{1}{2}\left(1/T_1^\mathrm{qb} + 1/T_1^\mathrm{TLS}\right) + 
\frac{1}{2}\left(\Gamma_{\pm} + \Gamma_{\mp} \right), \label{eq:T1} 
\end{equation}
where $\Gamma_{\pm}$ ($\Gamma_{\mp}$)  represents relaxation (excitation) between the two energy eigenstates $\ket{\pm} = (\ket{0g}\pm \ket{1e})/\sqrt{2}$ of \EqRef{eq:Hsub}, with energy splitting $h\fosc$. The polarization rate $\Gamma_{\pm} + \Gamma_{\mp} = S_{\bot}(\fosc)/2$ depends on the noise power $S_{\bot}$ that couples transversely to the diagonalized subspace Hamiltonian in \EqRef{eq:Hsub}, which for $\dph = 0$ corresponds to fluctuations $S_{\df}(\fosc)$ in the frequency detuning $\df$ \cite{Ithier:2005}.  Using \EqRef{eq:T1} and the measured values of $\tilde{T_1}$, $T_1^\mathrm{qb}$ and $T_1^\mathrm{TLS}$, we get $S_{\df}(f=76\MHz) = 2.8\times10^{6}\,\mathrm{rad/s}$.  We can not distinguish whether this noise comes from fluctuations in $\fqb$, $\ftls$ or a combination thereof, but we note that the measured value is a few times larger than fluctuations in $\fqb$ expected from flux noise. From independent measurements of the flux noise power $S_{\PhiQ}$ in the same device, we have $S_{\fqb} = S_{\PhiQ} (\dd{\fqb}{\PhiQ})^2 = 1.1\times10^{6}\,\mathrm{rad/s}$ at $f=76\MHz$ and $\PhiQ = -4.15\mPh$ \cite{Yan:2012}.

%% fit traces
%Figure~\ref{fig:vsN}(a) shows the dephasing rates $\Tph{0}$ and $\Tph{1}$, extracted by fitting the envelope of the coherent oscillations to \EqRef{eq:Decay}.  
%Both dephasing rates scales linearly with the flux detuning $\dph$, thus confirming that flux noise is the most likely origin of the dephasing. Also, note that the dephasing rate $\Tph{0}$ is 4-5 times stronger than $\Tph{1}$, consistent with a $1/f$-type spectrum \cite{Ithier:2005}.
%The scaling factor $\left|\dd{\fosc}{\Phi}\right|$ is essentially linear in $\dph$ for the range considered here.

\begin{figure}[tb]
\centering
\includegraphics[width=\linewidth]{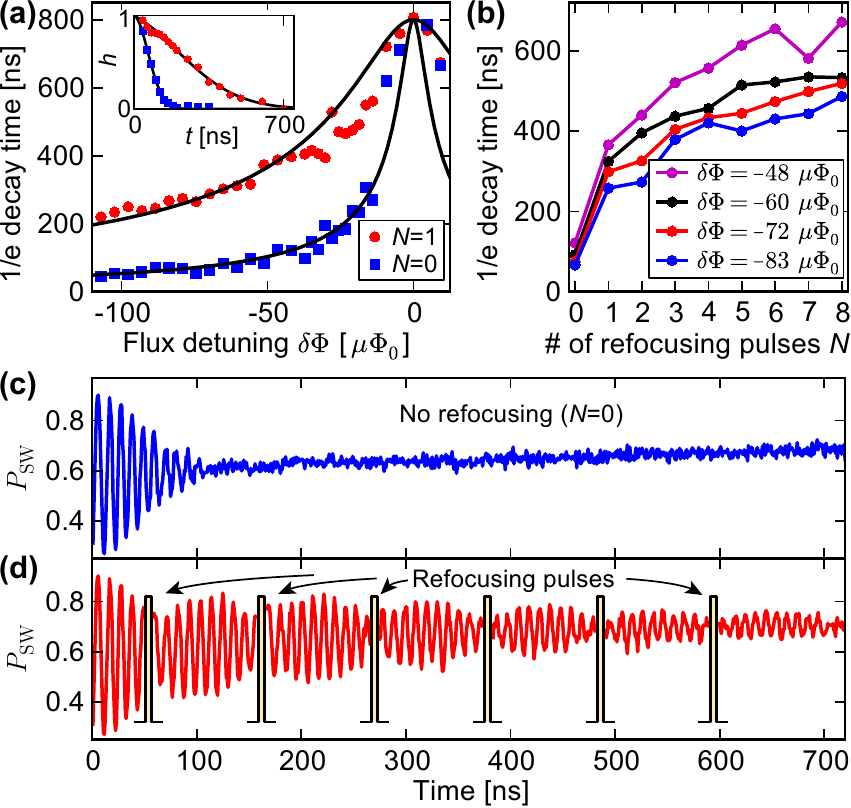}
\caption{(a) Decay times of the coherent oscillations, measured with ($N$=1) and without ($N$=0) a refocusing pulse.  
At $\dph=0$, the decay is limited by energy relaxation, but the coherence times decrease away from $\dph=0$ due to increased sensitivity to flux noise.  For large $|\dph|$, the refocusing sequence improves the decay time by more than a factor of four. The inset show examples of decay envelope $h(t)$, measured at $\dph = -60 \uPh$.  For the $N$=1 case, we use $\tau_1 = \tau_2 = t/2$.
(b) Decay times of multi-pulse refocusing sequences, showing the improvement coherence as the number of pulses increases. 
(c-d)  Time evolution of the qubit-TLS system, measured with and without refocusing pulses.  Note the echo signals appearing after each refocusing pulse, giving a strong enhancement of the coherence time. The data is taken at $\dph = -84 \uPh$.
}
\label{fig:vsN}
\end{figure}

%Having determined the decoherence due to relaxation, we focus on the pure dephasing part.  
Away from $\dph =0$, the decay envelope becomes Gaussian, and we extract the dephasing time $\TphN$ by fitting the data to \EqRef{eq:Decay}, assuming a constant relaxation time $\tilde{T_1} = 800\ns$.  
The extracted decay times versus flux $\dph$ are shown in \FigRef{fig:vsN}(a).  The refocusing sequence gives considerably longer decay times over the full range of the measurement except around $\dph=0$, where the decay is limited by $\tilde{T_1}$.  Note that to capture both the exponential and the Gaussian decay, we plot the time $T_e$ for the envelope to decrease by a factor $1/e$.  Examples of decay envelopes together with fits are shown in the inset of \FigRef{fig:vsN}(a), measured with and without a refocusing pulse at $\dph = -60\uPh$. 

We model the decreased phase coherence away from $\dph=0$ in terms of flux noise. In analogy with coherence measurements on single flux qubits \cite{Yoshihara:2006}, we assume $1/f$-type fluctuations in $\Phi$, with noise spectrum $S_\Phi(\omega)= A_\Phi/|\omega|$.  The flux noise couples to the oscillation frequency $\fosc$ through \EqRef{eq:fosc}, leading to the dephasing rate
\begin{equation}
 1/\TphN= 2\pi\sqrt{c_N A_\Phi}  \,  \big| \dd{\fosc}{\Phi}\big|. \label{eq:1f}
% \TphN = \sqrt{A_\Phi c_N}/(\hbar \Phi_0)  \left| \dd{\fosc}{\dph}\right|,
% \TphN = \frac{\sqrt{A_\Phi c^N}}{\hbar \Phi_0}  \left| \ddf{\fosc}{\dph}\right|,~~
% \Tph{0} = \frac{\sqrt{A_\Phi \ln(1/\omega_\mathrm{low} t)}}{\hbar \Phi_0}  \left| \ddf{\fosc}{\dph}\right|,~~
% \Tph{1} = \frac{\sqrt{A_\Phi \ln(2)}}{\hbar \Phi_0}  \left| \ddf{\fosc}{\dph}\right|
\end{equation}
Here, $c_{N=0} = \ln(1/\omega_\mathrm{low}t)$ and $c_{N=1}=\ln(2)$ relate to the filtering properties of the pulse sequence \cite{Ithier:2005, Bylander:2011}, with the low-frequency cut-off $\omega_\mathrm{low}/2\pi = 1\Hz$  fixed by the measurement protocol. 
% fit results
The solid lines in \FigRef{fig:vsN}(a) are fits to \EqRef{eq:1f}, using a single fitting parameter $A_\Phi = (1.4\uPh)^2$.  This amount of flux noise is consistent with previous results \cite{Yoshihara:2006,Bylander:2011}.
% zero flux detuning

%At $\dph=0$, $\fosc$ is first-order insensitive to changes in $\dph$, and the system is effectively decoupled from flux noise.

% discussion about method
The overall good agreement between \EqRef{eq:1f} and the data verifies the noise model and further confirms the validity of the refocusing sequence.  
% non-perfect results close to \dph=0
However, for the range $\dph>-30\uPh$, the refocused data shows slightly lower coherence times than expected from the model.
%In addition, the start to shows unexpected oscillations as a function of $\tau_1=\tau_2$.
We attribute this to the finite rise time of our shift pulses.
% adiabatic vs nonadiabatic
The refocusing sequence requires the frequency sweep rate $\dd{f}{t}$ to be fast compared to the interaction timescale $1/S\!\sim\!10\ns$ (to make the shifts non-adiabatic), but slow compared to the qubit precession time $1/\fqb\!\sim\!0.2\ns$ (to avoid driving the system out of the $\{\ket{1g},\,\ket{0e}\}$ subspace). The constraints can be phrased in terms of the probability of undergoing Landau-Zener transitions, giving $S^2 \ll \dd{f}{t} \ll \fqb^2$ \cite{Berns:2008}.
Using pulses with $1.5\ns$ Gaussian rise time,  we have $\dd{f}{t} \approx 550\MHz/1.5\ns = (606 \MHz)^2$, and on average the constraints are well fulfilled.  However, $\dd{f}{t}$ is lower during the slowest parts of the shift pulse (the beginning and the end), and artifacts due to imperfect non-adiabaticity appear when these parts of the pulse occur where the effective coupling is the strongest (at $\dph=0$).
%However, artifacts due to imperfect non-adiabaticity appear when the slowest parts of the shift pulse (the beginning and the end) occur where the effective coupling is the strongest (at $\dph=0$).  
The limited non-adiabaticity is also the reason for the slight asymmetry around $\dph = 0$ in \FigRef{fig:Setup}(c) \cite{Hofheinz:2009}.

% multi-pulse sequence
We now extend the refocusing technique to implement dynamical decoupling protocols with multi-pulse sequences.  For $1/f$-type noise, it has been shown that the Carr-Purcell sequence \cite{Carr:1954}, consisting of equally spaced $\pi$-rotations, improves coherence times by filtering the noise at low frequencies \cite{Faoro:2004, Falci:2004, Biercuk:2009, Bylander:2011}.  Figures~\ref{fig:vsN}(c-d) show the coherent evolution of the system when repeatedly applying refocusing pulses.  Echo signals form between each pair of $\pi$ pulses, giving considerable longer coherence times compared to the $N=0$ case.
The increase in decay time with the number of refocusing pulses $N$ is plotted in \FigRef{fig:vsN}(b), measured for a few different values of $\dph$.
%Figure~\ref{fig:vsN}(b) shows how the decay time increase with number of refocusing pulses $N$, measured for a few different values of $\dph$. 
The improvement is consistent with the filtering properties of the pulse sequence;  for larger $N$, the filter cut-off frequency increases and, since the noise is of $1/f$-type, the total noise power leading to dephasing is reduced.
%

% conclusion
To summarize, we have implemented refocusing and dynamical decoupling techniques to correct for noise and improve the lifetime of entangled two-level systems.  Although implemented between a flux qubit and a microscopic two-level system, the method applies to any transversely coupled spin-1/2 systems where the relative frequency detuning can be controlled.  We expect the findings to be of importance when developing decoupling techniques to improve two-qubit gate fidelities.

% samples

%\begin{acknowledgments}
We thank K. Harrabi for assistance with device fabrication and O. Zwier, X. Jin and E. Paladino for helpful discussions.  This work was supported by NICT. %Commissioned Research.
%\end{acknowledgments}

\bibliographystyle{apsrev}
\bibliography{RefocusedCoupling}
%@article{Yan:2012,
%author = {Fei Yan}, 
%journal = {in preparation},
%title = {in preparation},
%year = {2012},
%date-added = {2012-03-14 09:22:07 -0400},
%date-modified = {2012-03-14 09:23:16 -0400},
%uri = {papers://75EE009B-B603-4535-B1BD-919F5D5E42E3/Paper/p7563},
%rating = {0}
%}

\end{document}